\documentstyle[12pt,epsf]{article}

\def\hybrid{\topmargin -20pt    \oddsidemargin 0pt
        \headheight 0pt \headsep 0pt
        \textwidth 6.25in       
        \textheight 9.5in       
        \marginparwidth .875in
        \parskip 5pt plus 1pt   \jot = 1.5ex}

\hybrid

\def\baselinestretch{1.2}

\catcode`\@=11

\def\marginnote#1{}
\newcount\hour
\newcount\minute
\newtoks\amorpm
\hour=\time\divide\hour by60
\minute=\time{\multiply\hour by60 \global\advance\minute by-\hour}
\edef\standardtime{{\ifnum\hour<12 \global\amorpm={am}%
        \else\global\amorpm={pm}\advance\hour by-12 \fi
        \ifnum\hour=0 \hour=12 \fi
        \number\hour:\ifnum\minute<10 0\fi\number\minute\the\amorpm}}
\edef\militarytime{\number\hour:\ifnum\minute<10 0\fi\number\minute}

\def\draftlabel#1{{\@bsphack\if@filesw {\let\thepage\relax
   \xdef\@gtempa{\write\@auxout{\string
      \newlabel{#1}{{\@currentlabel}{\thepage}}}}}\@gtempa
   \if@nobreak \ifvmode\nobreak\fi\fi\fi\@esphack}
        \gdef\@eqnlabel{#1}}
\def\@eqnlabel{}
\def\@vacuum{}
\def\draftmarginnote#1{\marginpar{\raggedright\scriptsize\tt#1}}

\def\draft{\oddsidemargin -.5truein
        \def\@oddfoot{\sl preliminary draft \hfil
        \rm\thepage\hfil\sl\today\quad\militarytime}
        \let\@evenfoot\@oddfoot \overfullrule 3pt
        \let\label=\draftlabel
        \let\marginnote=\draftmarginnote
   \def\@eqnnum{(\theequation)\rlap{\kern\marginparsep\tt\@eqnlabel}%
\global\let\@eqnlabel\@vacuum}  }

\def\preprint{\twocolumn\sloppy\flushbottom\parindent 2em
        \leftmargini 2em\leftmarginv .5em\leftmarginvi .5em
        \oddsidemargin -.5in    \evensidemargin -.5in
        \columnsep .4in \footheight 0pt
        \textwidth 10.in        \topmargin  -.4in
        \headheight 12pt \topskip .4in
        \textheight 6.9in \footskip 0pt
        \def\@oddhead{\thepage\hfil\addtocounter{page}{1}\thepage}
        \let\@evenhead\@oddhead \def\@oddfoot{} \def\@evenfoot{} }

\def\numberbysection{\@addtoreset{equation}{section}
        \def\theequation{\thesection.\arabic{equation}}}

\def\underline#1{\relax\ifmmode\@@underline#1\else
        $\@@underline{\hbox{#1}}$\relax\fi}

\def\titlepage{\@restonecolfalse\if@twocolumn\@restonecoltrue\onecolumn
     \else \newpage \fi \thispagestyle{empty}\c@page\z@
        \def\thefootnote{\fnsymbol{footnote}} }

\def\endtitlepage{\if@restonecol\twocolumn \else \newpage \fi
        \def\thefootnote{\arabic{footnote}}
        \setcounter{footnote}{0}}  

\catcode`@=12
\relax

\def\figcap{\section*{Figure Captions\markboth
        {FIGURECAPTIONS}{FIGURECAPTIONS}}\list
        {Figure \arabic{enumi}:\hfill}{\settowidth\labelwidth{Figure
999:}
        \leftmargin\labelwidth
        \advance\leftmargin\labelsep\usecounter{enumi}}}
 \relax
\def\tablecap{\section*{Table Captions\markboth
        {TABLECAPTIONS}{TABLECAPTIONS}}\list
        {Table \arabic{enumi}:\hfill}{\settowidth\labelwidth{Table
999:}
        \leftmargin\labelwidth
        \advance\leftmargin\labelsep\usecounter{enumi}}}
 \relax
\def\reflist{\section*{References\markboth
        {REFLIST}{REFLIST}}\list
        {[\arabic{enumi}]\hfill}{\settowidth\labelwidth{[999]}
        \leftmargin\labelwidth
        \advance\leftmargin\labelsep\usecounter{enumi}}}
 \relax
\makeatletter
\newcounter{pubctr}
\def\publist{\@ifnextchar[{\@publist}{\@@publist}}
\def\@publist[#1]{\list
        {[\arabic{pubctr}]\hfill}{\settowidth\labelwidth{[999]}
        \leftmargin\labelwidth
        \advance\leftmargin\labelsep
        \@nmbrlisttrue\def\@listctr{pubctr}
        \setcounter{pubctr}{#1}\addtocounter{pubctr}{-1}}}
\def\@@publist{\list
        {[\arabic{pubctr}]\hfill}{\settowidth\labelwidth{[999]}
        \leftmargin\labelwidth
        \advance\leftmargin\labelsep
        \@nmbrlisttrue\def\@listctr{pubctr}}}
 \relax
\makeatother
\newskip\humongous \humongous=0pt plus 1000pt minus 1000pt

\newif\ifdtup

\relax

\def\be{\begin{equation}}
\def\ee{\end{equation}}
\def\ba{\begin{eqnarray}}
\def\ea{\end{eqnarray}}

\def\m{\mu}
\def\n{\nu}

\def\no{\noindent}

\def\IR{\relax{\rm I\kern-.18em R}}

\def\IR{\relax{\rm I\kern-.18em R}}
\def\inv{^{\raise.15ex\hbox{${\scriptscriptstyle -}$}\kern-.05em 1}}

\def\tL{{\tilde L}}

\begin{document}

\renewcommand{\theequation}{\arabic{equation}}

\newcommand{\beq}{\begin{equation}}
\newcommand{\eeq}[1]{\label{#1}\end{equation}}
\newcommand{\ber}{\begin{eqnarray}}
\newcommand{\eer}[1]{\label{#1}\end{eqnarray}}
\newcommand{\eqn}[1]{(\ref{#1})}
\begin{titlepage}
\begin{center}

\hfill CERN-TH/99-192\\
\hfill hep--th/9906204\\

\vskip 1.2in

{\large \bf Exponential and Power-Law Hierarchies from
 Supergravity}

\vskip 0.6in

{\bf A. Kehagias}
\vskip 0.1in
{\em Theory Division, CERN\\
     CH-1211 Geneva 23, Switzerland\\
{\tt kehagias@mail.cern.ch}}\\
\vskip .2in

\end{center}

\vskip .6in

\centerline{\bf Abstract }

\no
We examine how a $d$-dimensional  mass hierarchy  can be generated
from a $d+1$-dimensional set up. We consider   a 
$d\!+\!1$--dimensional scalar, the hierarchon, which has a potential as in 
 gauged supergravities.  We find that when  it is in its minimum, 
there exist solutions of Ho$\check{r}$ava-Witten topology
$R^{d}\times S^1/{\bf Z^2}$  with domain walls at the 
fixed points and anti-de Sitter geometry in the bulk.
We show that while  standard Poincar\'e 
supergravity leads to  power-law hierarchies, (e.g.   
a power law dependence of masses  on the compactification scale), 
gauged supergravity produce an exponential hierarchy as recently proposed 
by Randall and Sundrum.

\vskip 0,2cm
\no

\vskip 4cm
\noindent
CERN-TH/99-192\\
June 1999\\
\end{titlepage}
\vfill
\eject

\def\baselinestretch{1.2}
\baselineskip 16 pt
\noindent

\def\tT{{\tilde T}}
\def\tg{{\tilde g}}
\def\tL{{\tilde L}}


\section{Introduction}

A puzzling feature in all efforts to extend the Standard Model (SM) is 
the hierarchy $m_{EW}/M_{Pl}$ of the electroweak scale 
$m_{EW}\sim 10^3 \ {\rm GeV}$ and the Planck scale $M_{Pl}= G_N^{1/2}\sim
10^{18}\ {\rm GeV}$.  
 A  proposal for explaining 
this hierarchy has been made in \cite{ADD} realizing recent ideas on the size 
 of the compactification, the string scale and the coupling constants 
\cite{W}. According to this 
proposal, the higher $4+n$ dimensional theory with $n\geq 2$ has a $4+n$ 
dimensional Planck mass $M_{Pl(4+n)}$ at the TeV scale 
while the scale $R_c$ of the extra $n$ dimensions is less than a millimeter. 
The proposal has been designed in such a way as to generate  the hierarchy  
$m_{EW}/M_{Pl}$. However, it suffers from another hierarchy,
that of $m_{EW}R_c$. 

In \cite{RS},  an alternative scenario was proposed for generating the 
hierarchy  without large extra dimensions. According to this, 
the four-dimensional metric is not factorizable but rather is multiplied 
by a warp factor with exponential dependence on a transverse coordinate which 
has finite range. The overall space is in fact a portion of a 
five-dimensional anti-de Sitter space-time (AdS). 
Masses of the four-dimensional world  
are also multiplied with this warping factor which can generate hierarchies 
for not neseccarily compactification radius.     

Along these lines, we will try to push forward the idea that mass hierarchies 
in four dimensions can be smoothed out in a higher-dimensional 
setting. We will show that exponential hierarchy can actually be produced by 
a scalar in five dimensions with a potential like the one  in gauged 
supergravities. We will call this scalar {\it hierarchon} 
for obvious reasons. In fact it is one of the scalars which appear in gauged
supergravities and  have recently been discussed in connection with the 
renormalization group flow \cite{ZZ,KPW,FGPW} 
in the context of AdS/CFT correspondence \cite{Mald,KP,WW}.
Recalling that according to the general belief, solutions of 
gauged supergravities are true compactifications of type IIB or 
eleven-dimensional supergravity,\footnote{For the four-dimensional $SO(8)$ 
gauged supergravity this has been proven in \cite{NW}.} our solutions 
should also corresponds to ten or eleven-dimensional 
supersymmetric backgrounds.  

The potential
of the hierarchon is such that it has negative value at its minimum and 
effectively produce a cosmological constant. In this case, there 
are solutions which describe flat domain walls separated by a bulk 
AdS geometry and correspond to an $S^1/{\bf Z}_2$ compactification of the 
radial AdS coordinate. Similar solutions in the context of M-theory have
been found in \cite{LS1}. 
An induced exponential hierarchy on the walls is then 
generated as
in \cite{RS} which depence on the value of the potential at its 
critical points. In the examples we study, exponential hierarchy is generated
on one of the boundaries only while on the other we find a power-law 
hierarchy.  Finally, when there is no such scalar as in the usual ungauged
Poincar\'e supergravity, the hierarchy is always power-law and a scenario 
with large extra dimensions should be taken as in \cite{ADD}.

\section{Exponential hierarchy in gauged supergravity}

We will start by recalling some results from  gauged-supergravity theories.
These theories exist only in 
$d=3,...,7$ dimensions and they are ultimate related to the existence 
of AdS supergroups in these dimensions \cite{N}.
For their construction,
one usual starts with the  ungauged d-dimensional 
Poincar\'e supergravity  which, generally, contains a 
set of scalars  parametrizing a coset $G/H$ and a set of (abelian)
vectors $A_M^I$ transforming in a specific representation ${\bf r}$ of $G$. 
The gauged supergravity 
is then  obtained from the ungauged one by gauging an appropriate non-abelian 
subgroup  $K\subset G$. The gauge group $K$ is such that  
the decomposition of ${\bf r}$ in representations of $K$ 
contains the adjoint. The construction proceeds by replacing the abelian 
vectors with non-abelian ones. This replacement
clearly violates supersymmetry which, however, can be recovered by 
adding terms in the action and 
changing appropriately the supersymmetry transformation rules. 
This procedure generates a potential for the scalar fields with non-trivial
critical points   which is the 
characteristic of gauged supergravities. 

The effective action of the $d=4,7$ and $d=5$  gauged supergravities can 
be obtained by KK compactification of eleven-dimensional supergravity and 
ten-dimensional type IIB supergravity on $S^4,S^7$ and $S^5$, respectively.
Since we are mainly interested in the $d=5$ case, let us describe it in more 
details. 

The  toroidal compactified  type IIB string
theory in five dimensions has  global $E_{6(+6)}$ and a local $ USp(8)$ 
symmetry. The massless spectrum fills the ${\cal N}=8$ five-dimensional 
graviton multiplet which consists of a graviton, 8 gravitini, 27 vectors, 48
gauginos and 42 scalars \cite{CR}.  The  scalars parametrize the non-compact 
symmetric space $E_{6(+6)}/USp(8)$. An $SO(6)=SU(4)$ subgroup of  $E_{6(+6)}$
can be gauged leading to the ${\cal N}=8$ gauged supergravity \cite{GRW,GRW1}.
The potential of the latter is rather involved, it is $SU(4)$-invariant  
and all the $42$ scalars 
appear  in it except the dilaton and the axion. There is no classification 
of the critical points of the potential although some of the vacua are known 
\cite{KPW,FGPW,GRW,GRW1}. 

Turning to the general case, the symmetry group $K$ can be broken to a 
subgroup $K_0\subset K$ by the expectation value of a scalar $\lambda$ 
which is a $K_0$ singlet. We will call the scalar $\lambda$ which takes an 
expectation value  {\it hierarchon}
since, as we will see, it will set up the hierarchy.      
The bosonic part of the gauged supergravity action for the 
graviton and the hierarchon is then of the form
\be
S=\frac{1}{2\kappa_d^2}\int d^dx\sqrt{G} \left(\phantom{\frac{Y}{X}}
\!\!\!\!\!\!\!
R-\frac{1}{2}\partial_M\lambda\partial^M\lambda-V(\lambda)\right)\ , 
\label{EH}
\ee
where $\kappa_d^2$ is the gravitational constant in $d$-dimensions. 
The potential $V(\phi)$ 
has, generically, critical points 
for $\lambda=\lambda_0=const$ and the value of $V$ at these points 
can conveniently be parametrized as $V(\phi_0)=-\frac{(d-2)(d-1)}{L^2}$.
For example, in the case of five-dimensional gauged supergravity, the
potential for the $SO(5)$-invariant hierarchon is 
\be
V(\lambda)=-\frac{1}{32}g^2\left(15e^{2\lambda}+10e^{-4\lambda}-e^{-10\lambda}
\right)\, , 
\ee
where $g$ is the gauge-coupling constant. The above potential has 
an $SU(4)$ symmetric minimum at $\lambda_0=0$ where $V(\lambda_0)=-3g^2/4$
with ${\cal N}=8$ supersymmetry.
There is also an other minimum at $\lambda_0=-1/6\log3$ with 
$V(\lambda_0)=-3^{5/3}g^2/8$ with  $SO(5)$ symmetry and no supersymmetry, 
${\cal N}=0$. Similarly, if $\lambda$ is taken to be an $SU(3)\times U(1)$
singlet, the potential turns out to be 
\be
V(\lambda)=\frac{3}{32}g^2\left(\cosh(4\lambda)^4-4\cosh(4\lambda)-5\right)\, .
\ee
Here, the $SU(3)\times U(1)$-symmetric minimum exist at $\lambda_0
=1/4\cosh^{-1}2$ with $V(\lambda_0)=-27/32g^2$ \cite{GRW2}. The above 
potentials have been employed in \cite{ZZ} for the discussion of the 
renormalization-group flow in the $AdS/CFT$ correspondence. 

Having frozen the hierarchon field at its minimum, 
the rest of the field 
equations are simply 
\be
R_{MN}= - \frac{d-1}{L^2}\ G_{MN}\ . \label{fe}
\ee  
The supersymmetry transformations for the  gravitino take the form 
\be
\delta \psi_M=\partial_M \epsilon - \frac{1}{2L}i
\Gamma_M\epsilon\, .
\label{gravitino}
\ee
The obvious solution to the above equations is the anti-de Sitter space 
$AdS_{d}$ which is the unique maximally symmetric space with negative 
cosmological constant. It is found as the vacuum solutions of 
gauged supergravity for $d=7,5,4$ and describe M-theory vacua on $AdS_{4,7}
\times S^{7,4}$ and type IIB on $AdS_5\times S^5$. They are also realized as
the near-horizon limit of various brane-configurations \cite{GT} 
and they are the 
supergravity duals of superconformal field theories \cite{Mald,KP,WW}.  
The isometry group of  
$AdS_d$ spaces are  $SO(2,d-1)$ which is the conformal group in $d$--$1$ 
dimensions.  However, here we will 
describe solutions to the field equations \eqn{fe} 
which are not maximally symmetric but rather invariant under 
the group $ISO(1,d-2)\times U(1)$ where $ISO(1,d-2)$ is the Poincar\'e group
in $d-1$ dimensions.       
The ansatz for the $d$-dimensional metric is then of the form 
\be
ds^2=H(z)^{2a}\eta_{\mu\nu}dx^\m dx^\n +H(z)^{2b}dz^2\ ,
\label{mm}
\ee
where $H(z)$ is a function of the transverse coordinate $z$. The constants
$a,b$ can be  determined by  demanding supersymmetry, 
that is the solution 
to be annihilated by the supercharges. This is equivalent to the vanishing 
of  all 
fermionic shifts. 
The integrability condition  
$\delta\psi_M=0$ turns out to be
\be
R_{MNAB}\Gamma^{AB}\epsilon=\frac{2}{L^2}\Gamma_{MN}\epsilon \, . 
\label{integr}
\ee
For the metric in \eqn{mm}, the integrability condition 
gives that  $a^2=1, b=-1$. Moreover, $H(z)$
is a harmonic function in the transverse z-direction, i.e., it satisfies
\be
H''=0\, ~~~~ {\mbox {with}}~~~~ {H'}^2=1/L^2\, , \label{HH}
\ee 
where the prime $(')$ denotes differentiation $d/dz$. For $a=1$
the solution turns out then to be
\ba
ds^2&=&H^{2}\eta_{\mu\nu}dx^\m dx^\n +H^{-2}dz^2\ ,
\label{metric} \\
H&=& \frac{1}{L}z+c\, . \label{h}
\ea
We will not consider 
the case $a=-1$ since it is a coordinate transformation of the $a=1$ case.
Clearly the metric  above describes a space-time 
invariant under the Poincar\'e group 
$ISO(1,d-2)$ in the longitudinal $d\!-\!1$ dimensions. In fact, if $H(z)$ is 
continuous, the symmetry is $SO(2,d-1)$ since the metric \eqn{metric}
describes a d-dimensional anti-de Sitter space. On the other hand, if 
$H(z)$ is piecewise continuous, it describes domain walls sited at the 
discontinuous  points. Such solutions have been discussed in 
\cite{T,LPT,T1,GR,C}.
Another possibility is to consider the case in which, $H(z)$ is piecewise 
continuous and periodic \footnote{We thank L. Alvarez-Gaum\'e for pointing
out this possibility}.  Among many possibilities, we will discuss  
two particular cases 
as depicted in  figure 1 which solves \eqn{HH}. 

\begin{figure}[htb]
\epsfxsize=4in
\bigskip
\centerline{\epsffile{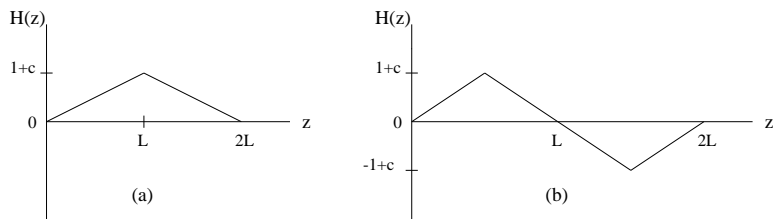}}
\caption{The function $H(z)$ for $0\leq z<2L$.}
\bigskip
\label{fig1}
\end{figure}

We will examine the two cases (a) and (b) separately.
\vspace{.2cm}

\noindent
{\bf The (a) case}
\vspace{.1cm}

Here, 
$H(z)$ is of the form
\be
H(z)=\left\{ \begin{array}{ll}
\phantom{-}\frac{1}{L}z + c &\phantom{XXXX} 0\leq z<L  \\
-\frac{1}{L}z + c & \phantom{XXXX}L\leq z<2L \end{array}\right. \label{h1}
\ee
$H(z)$ is discontinuous but nevertheless it has the right properties, namely
it is  periodic with period $2L$ as can easily be seen in its Fourier-series 
representation
\be
H(z)=\frac{1}{2}+c -\frac{4}{\pi^2}\sum_{n=1,3,5,\cdots}\frac{1}{n^2}
\cos\left(\frac{n\pi z}{L}\right)\, , \label{HF}
\ee
and satisfies the second equation in \eqn{HH} everywhere while the 
first in \eqn{HH} ``almost'' everywhere. 
By recalling the series representation of the $\delta$-function 
\be
\delta(z)=\frac{1}{2L}+\frac{1}{L}\sum_{n=1}^{\infty}
\cos\left(\frac{n\pi z}{L}\right)\, , \label{delta}
\ee
we find that 
\be
H''=\frac{2}{L}\left(\!\!\!\!\!\!\phantom{\frac{X}{X}}
\delta(z)-\delta(z+L)\right)\, , \label{hjk}
\ee
and thus the discontinuity of $H(z)$ can be attributed to sources at the 
discontinuous points. The form of the sources can be found by 
considering the energy-momentum tensor $T_{MN}$ which in our case is 
given by 
\be
T_{MN}=
\frac{1}{\kappa_d^2}\left(R_{MN}-\frac{1}{2}G_{MN}R-
\frac{(d-2)(d-1)}{2L^2}G_{MN}\right)\, .
\ee
Then,  it is straightforward to verify that for the metric \eqn{metric}
the energy-momentum tensor is given by
\ba
T_{\m\n}&=&\frac{d\!-\!2}{\kappa_d^2}\ H''\ H^3\
\eta_{\m\n}\ , ~~~~~T_{zz}=0\, ,
\ea
so that, by using  \eqn{hjk} 
\be
T_{\m\n}=\frac{2(d\!-\!2)}{L\kappa_d^2}
\left(\!\!\!\!\!\!\phantom{\frac{X}{X}}
\delta(z)-\delta(z+L)\right)\ H^{3}\
\eta_{\m\n}\, .
\ee
Thus, our solution describes two domain walls placed at 
$z=0$ and $z=L$, respectively. In particular,  the  solution is 
invariant under $z\to 2L-z$ as can be seen from  \eqn{HF} and   there 
exist two fixed points, the $z=0$ and $z=L$. These  are the points where 
our domain walls are sited. In this case we may restrict $z$ to be in the 
interval
$[0,L]$ corresponding to an $S^1/{\bf Z}_2$ orbifold of the transverse 
one-dimensional space. A similar solution has also been found in 
\cite{LS1} in the context of  
M-theory where  $z$ is identified with the eleventh dimension.

The constant $c$ in $H(z)$ is determined in terms of the compactification
radius $R_c$ and the cosmological constant $L$. From \eqn{metric}
we see that the compactification radius is
\be
R_c=\frac{1}{\pi}\int_0^L H^{-1}dz = \frac{L}{\pi}\ln\left(1+\frac{1}{c}
\right)\, ,
\ee
so that  we get
\be
c=\left(e^{\pi R_c/L}-1\right)^{-1}
\ee  

Let us now suppose that  gauge theories live on the domain walls found above
while gravity propagates in the bulk. Then for the  case 
$(a)$ in figure 1,  
we see from \eqn{metric} 
that the  masses $m$ in the $d$-dimensional theory 
as measured from the domain wall flat metric $\eta_{\m\n}$ at 
$z=0$ and $z=L$, respectively, are 
\ba
m_0&=&H(0)m=\left(e^{\pi R_c/L}-1\right)^{-1}m \, ,
\label{m0}\\
m_L&=&H(L)m_0=
\left(\frac{e^{\pi R_c/L}}{e^{\pi R_c/L}-1}\right)^{-1}m \, , 
\ea
 since $H(0)=c$ and $ H(L)=c+1$. For 
$R_c>L/\pi$ we find  that
\be
m_0\approx e^{-\frac{\pi R_c}{L}}m \, , ~~~~m_L\approx m \, . 
\ee
Thus, while at $z=L$ masses as measured in the full $d$-dimensional 
metric and the wall flat metric are of the same order, they are exponentially
suppressed at $z=0$. As a result, we get exponential hierarchy in one 
boundary as in \cite{RS}.
\vspace{.2cm}

\noindent
{\bf The (b) case}
\vspace{.1cm}

Let us now discuss the second example  $(b)$ in figure 1 for the function 
$H(z)$. Here, $H(z)$ is 
\be
H(z)=c+\frac{8}{\pi^2}\sum_{n=1,3,5,\cdots}\frac{(-1)^{(n-1)/2}}{n^2}
\sin\left(\frac{n\pi z}{L}\right)\, , \label{HFb}
\ee
which is again periodic with period $2L$. By using \eqn{delta},
it is straightforward to verify that
\be
H''=\frac{4}{L}\left(\!\!\!\!\!\!\phantom{\frac{X}{X}}
\delta(z+\frac{L}{2})-\delta(z+\frac{3L}{2})\right)\, ,
\ee
and that the energy-momentum tensor is in this case
\be
T_{\m\n}=\frac{4(d\!-\!2)}{L\kappa_d^2}
\left(\!\!\!\!\!\!\phantom{\frac{X}{X}}
\delta(z+\frac{L}{2})-\delta(z+\frac{3L}{2})\right)\ H^{3}\
\eta_{\m\n}\, .
\ee
Thus, again the solution \eqn{HFb} describes two domain walls placed 
at $z=L/2,3L/2$ respectively. The constant $c$ can 
also be determined in terms of the cosmological 
constant $\sim 1/L^2$ and the compactification radius $R_c$ which 
in this case is found to be
\be
R_c=\frac{L}{2\pi}\int_0^{2L}H^{-1}dz=\frac{L}{2\pi}\ln
\left(\frac{c+1}{c-1}\right)\, . \label{rr} 
\ee
Then we find from \eqn{rr} that $c$ is given in this case 
by
\be
c=\frac{e^{2\pi R_c/L}+1}{e^{2\pi R_c/L}-1}\, . \label{cc} 
\ee

Similarly to the case (a),  we find that masses 
$m_{L/2}$ and $m_{3L/2}$ measured with
the flat $\eta_{\m\n}$ metric at the domain walls in $z=L/2$ and 
$z=3L/2$, respectively are related to the mass $m$ measured with the metric 
\eqn{metric}
by
\be
m_{L/2}= \left(\frac{2e^{2\pi R_c/L}}{e^{2\pi R_c/L}-1}\right)^{-1}m
\, , ~~~~    
m_{3L/2}= \left(\frac{2}{e^{2\pi R_c/L}-1}\right)^{-1}m \, .
\ee
Thus, again for $R_c>L/2\pi$ we get
\be
m_{L/2}\approx m\, , ~~~~
m_{3L/2}\approx e^{-\frac{2\pi R_c}{L}}m \, ,  
\ee
so that we have exponential hierarchy in one of the boundaries.
This seems to suggest that  this type of behaviour, i.e., 
exponential hierarchy at one of the two boundaries is universal.   

\section{Power-law hierarchies in ungauged supergravity}

Let us now turn to standard supergravity in which there exist p-brane 
solutions involving the metric $G_{MN}$, the dilaton $\Phi$ and an 
antisymmetric form field strength of rang $n$. The p-brane is a
charged object with electric charge  if
$p=n-2$ and in this case  the p-brane is elementary.
If $p=d-n-2$, where $d$ is the space-time dimensions, the p-brane has magnetic 
charge and it  is 
solitonic. In the latter case we may also have $n=0$ which means that there 
is no antisymmetric field strength but rather a cosmological-type term 
The 
bosonic part of the supergravity action is then \cite{T,LPT}
\be
S=\frac{1}{2\kappa_d^2}\int d^dx \sqrt{G}\left(R-\frac{1}{2}
\partial_M\Phi\partial^M\Phi-\frac{2}{L^2\Delta}e^{-\alpha\Phi}\right)
\, , \label{ac}
\ee
where $\Delta=\alpha^2-2(d-1)/(d-2)$. The action \eqn{ac}
admits solitonic $d-2$-brane solution of the form
\ba
ds^2&=&H^{\frac{4}{\Delta(d-2)}}\eta_{\m\n}dx^\m dx^\n+ 
H^{\frac{4(d-1)}{\Delta(d-2)}}dz^2\, , \label{metric2} \\
e^{\Phi}&=&H^{\frac{2\alpha}{\Delta}} \, , \label{Phi}
\ea
where, again $H=H(z)$ satisfies \eqn{HH}. By choosing $H$ as in figure 1.
we get a background with two domain walls as in the gauged supergravity
discussed above. For the case (a),
the compactification radius $R_c$ turns out to be
\be
R_c=\int_0^L H^{\frac{2(d-1)}{\Delta(d-2)}}dz=
\frac{\Delta L}{\pi \alpha}\left((1+c)^{\alpha^2/\Delta}-c^{\alpha/\Delta}
\right)\, . \label{rp}
\ee
Then, the masses $m_{L/2}$ as measured with the boundary 
flat metric $\eta_{\m\n}$ are related to the masses $m$ 
measured with the  bulk metric \eqn{metric2} by $m_{L/2}=
c^{\frac{2}{\Delta(d-2)}}m$ and a hierarchy may be  generated for $c\ll1$.
From \eqn{rp} we find  then 
\be
m_{L/2}\approx \left(\frac{\pi R_c}{L}-\frac{\Delta}{\alpha^2}
\right)^{\frac{2}{\Delta(d-2)}}\ m
\ee
i.e., a power-law hierarchy. This is a general 
feature in this kind of solutions, namely, the $\alpha\neq$ case,
which corresponds to domain wall solution in standard Poincar\'e 
supergravity, always leads
to power-law hierarchy since the constant $c$ is a power of the 
compactification radius $R_c$. On the other hand, in gauged supergravities
we may have $\alpha=0$ leading to an exponential dependence of $c$ on 
$R_c$ and consequently to exponential hierarchies as explained in the 
previous section.

\vspace{.5cm}

\noindent 
{\bf Acknowledgement:} We would like to thank L. Alvarez-Gaum\'e, E. Kiritsis,
A. Riotto and A. Zaffaroni for discussions. 

\vspace{.2cm}

\noindent
{\bf Note added:} While this work was in its final stage, we received \cite{V}
where  backgrounds which realize exponential hierarchy was also constructed.

\end{document}